\documentclass[preprint,longnamesfirst]{aastex}

\newcommand{\thisdir}{.}

\newcommand{\cm}{\mathop{\rm cm\,}\nolimits}

\newcommand{\kpc}{\mathop{\rm kpc\,}\nolimits}
\newcommand{\Mpc}{\mathop{\rm Mpc\,}\nolimits}

\newcommand{\K}{\mathop{\rm K\,}\nolimits}
\newcommand{\g}{\mathop{\rm g\,}\nolimits}

\newcommand{\boltz}{\mathop{\rm k\,}\nolimits}
\newcommand{\G}{\mathop{\rm G\,}\nolimits}

\newcommand{\fig}{Fig.~\ref}

\newcommand{\tab}{Table~\ref}

\newcommand{\sect}{Sec.~\ref}

\newcommand{\eq}{Eq.~\ref}
\newcommand{\Hydra}{{\sc hydra}}

\newcommand{\expd}[1]{\times 10^{#1}}

\newcommand{\fraction}[2]{\mbox{\scriptsize$^{{#1}\!}/_{\!{#2}}$}}

\ifx\undefined\onequarter
 \newcommand{\onequarter}{\fraction{1}{4}}
\else
 \renewcommand{\onequarter}{\fraction{1}{4}}
\fi
\ifx\undefined\onethird
 \newcommand{\onethird}{\fraction{1}{3}}
\else
 \renewcommand{\onethird}{\fraction{1}{3}}
\fi
\ifx\undefined\onehalf
 \newcommand{\onehalf}{\fraction{1}{2}}
\else
 \renewcommand{\onehalf}{\fraction{1}{2}}
\fi
\ifx\undefined\twothirds
 \newcommand{\twothirds}{\fraction{2}{3}}
\else
 \renewcommand{\twothirds}{\fraction{2}{3}}
\fi
\ifx\undefined\threequarters
 \newcommand{\threequarters}{\fraction{3}{4}}
\else
 \renewcommand{\threequarters}{\fraction{3}{4}}
\fi

\newcommand{\figscale}{1}
\newcommand{\figscaletwo}{1}

\begin{document}

\title{RESOLVING THE JEANS MASS IN HYDRODYNAMIC SIMULATIONS OF HIERARCHICAL CLUSTERING}
\shorttitle{JEANS MASS IN HIERARCHICAL CLUSTERING}
\author{Eric R. Tittley\footnote{Present address: University of Maryland, Baltimore County, Department of Physics , Baltimore, MD, 21250}
 and 
 H. M. P. Couchman\footnote{Present address: McMaster University, Department of Physics and Astronomy, Hamilton, Ontario, L8S 4M1, Canada}}
\email{tittley@umbc.edu}
\email{couchman@physics.mcmaster.ca}
\affil{University of Western Ontario}
\affil{
Department of Physics and Astronomy, 
London, Ontario, N6A 3K7, 
Canada }
\shortauthors{TITTLEY \& COUCHMAN}

\begin{abstract} 
We examine the necessity of resolving the Jeans mass in
hydrodynamical simulations of the hierarchical formation of
cosmological structures. We consider a standard
two-component fluid and use smoothed particle hydrodynamics to model
the baryonic component. It is found that resolution of the Jeans mass
is not necessary to extract bulk properties of bound objects that are
independent of resolution, provided that the objects have undergone
a number of merger events. The degree of merging must be sufficient to
populate the objects with substructure of similar total mass.
After this criterion is met, neither the density profiles of
structures nor their density-temperature distributions depend
appreciably on the resolution of the simulation.  Resolution of the
Jeans mass by the imposition of a minimum temperature is not found to
adversely affect the state of the matter in the final clusters.  The
baryon concentration profiles still indicate a deficit of gas with
respect to the dark matter in the interiors of structures.
\end{abstract}


\section{INTRODUCTION}
\label{sec.Intro}
Numerical simulations have proved useful for understanding the history
and dynamics of cosmological structure
\citep{DEFW92,Katz92,Cen94,EMN}. Our ability to model collisionless
systems has matured to the extent that bulk properties of the dark
matter can be predicted with increasing confidence, although important
exceptions exist, such as the merger rates of substructure as well as
the inner profiles of dark matter halos \citep[see, for
example,]{Moore98}.

\shortcites{Frenk98}
Combining collisionless N-body techniques with a hydrodynamical method
to follow the baryonic component opens the way to a detailed
understanding of cosmic structure formation in a variety of
models. Modeling the baryonic component removes the need to make
assumptions about the relationship between the baryonic material and
the underlying collisionless matter. Such techniques have proved to be
both popular and powerful
\citep{HK89,Evrard90,Katz91,Cen92,Navarro93,SM93,ROKC93,Bryan94,Gnedin95,Steinmetz96}.
A recent comparison of several hydrodynamic codes \citep{Frenk98} has
shown that we may also begin to have confidence in the overall
properties of the baryonic distribution predicted by these codes.
However, several differences which are apparent both between codes and
between results of varying resolution suggest that convergence has yet
to be established conclusively.  We address here the particular issue
of convergence with resolution in hydrodynamic simulations in which
objects are built up through hierarchical merging. We do not consider
radiative cooling in this work.

In the standard model of cosmological structure formation, larger
structures are formed via the amalgamation of smaller structures
formed at earlier times in a hierarchical fashion.  In any simulation
which models the formation of such structures in this hierarchical
manner, there will be an abundance of objects formed at the resolution
limit of the simulation.  Numerical effects will be a particular
problem for these objects.  The propagation of these effects into the
larger objects subsequently formed requires characterization both to
determine the reliability of the physical model as well as
to ensure convergence among simulations performed at various resolutions.

Recently \citet{OV97} has argued that it is necessary to resolve the
Jeans mass in a simulation if convergence is to be attained with
increasing resolution. This requires that a minimum temperature be
imposed in the simulation such that the gas in the first bound
objects---which form at the resolution limit in a standard simulation
of hierarchical clustering---is at a sufficient temperature to support itself via pressure in the gravitational potential well of the first dark
matter structures.  They argue that if the gas temperature is
initially much less than the virial temperature of the first halos
(this is a boundary condition frequently adopted in cosmological
simulations), the cold gas falling into the first potential wells will
suffer spuriously large amounts of shock heating of the gas. This ``shocking''
will be greater for simulations with lower resolution since the scale
over which the shocks occur will be larger. Conversely, if the initial
generation of objects is pressure-supported, the interaction of the
baryons in merging halos as the hierarchical structure formation
proceeds will lead only to weak shocks. (This result follows since the
velocity with which lumps merge is similar to the internal velocity
dispersion of the individual lumps in a typical hierarchy.) Owen and
Villumsen suggests that the difference in the amount of shock heating
at early stages will persist through to the final object and prevent
convergence of numerical results unless a minimum temperature is
imposed.

\citeauthor{OV97} finds, in scale-independent simulations, that
without the imposition of a minimum temperature to pressure support
resolution-limited structures, the gas is found to move {\em en masse}
toward a state of higher density and lower temperature as the
resolution of the simulation increases. This behavior is not to be
expected during the formation of hydrostatic structures. In the
standard model, we expect the virial temperature of a structure to be
broadly determined by its mass, following the relationship $T \propto
M^{2/3}$.  Since the bulk of the gas resides in the halos of these
structures, the temperature of the bulk of the gas should consequently be
determined by the dark matter distribution, a distribution whose
convergence among differing resolutions is confirmed in the
simulations of \citeauthor{OV97}. 

There are other issues of concern raised by the results of the simulations described in \citet{OV97}.  First, the results suggest paradoxically that the injection of thermal energy into a system at an early time can lead to a cooler system at a later time.  The authors claim that this is achieved by reducing the amount of shock-heating as the first, unresolved, objects collapse.  However, the injection of thermal energy into the system via the imposition of a minimum temperature pre-heats the gas in a manner which is essentially equivalent to shock-heating the gas in the first objects.  

Another concern is the concentration of baryons relative to the dark matter in the cores of structures reported in \citeauthor{OV97}. In three-dimensional simulations without cooling, spanning a wide range of resolutions, others have found {\em extended} distributions of baryonic material, relative to the dark matter \citep{Evrard90,TC92,CO93a,Kang94,ME94,PTC94,NFW95,AN96,Lubin96,PEG96}. \citet{PTC94} explains the phenomenon as a result of the merging process in which gas is shocked, permanently removing energy from the dark matter component and passing it to the gas.

A possible explanation for the conflicting results regarding the baryon concentration may be that the simulations of \citeauthor{OV97} were done using a 2-D code.  This allowed them to achieve high spatial resolutions but may have introduced phenomena unique to two-dimensional geometries. 

\subsection{Jeans Mass in a Two-component Fluid}
In a self-gravitating baryonic fluid, the Jeans mass is easily obtained, by a perturbatitive analysis, as a sharp threshold between acoustic oscillation and collapse, depending on the density and temperature of the fluid. In a two-component fluid comprised of baryons and collisionless cold dark matter, the situation is more complicated. The low initial velocity dispersion of cold dark matter leads to instability, or growth, in this component on all scales. The relevant question in this case for the baryonic fluid is on what scales do the two components remain coupled adiabatically? On small scales, gas pressure will be sufficient to oppose the gas falling into the small dark matter potential wells, whereas on large scales the initial gas temperature will be lower than the effective virial temperature of the forming dark matter halo and the gas will increase in overdensity with the dark matter. A perturbative analysis of the coupled two-component fluid is relatively straightforward but not very illuminating for our purposes. In particular it can be shown that there is not a single Jeans mass in the sense described above \citep[see][for example]{Peacock99}. It is sufficient for our purposes to define the Jeans mass as that which is just sufficient to adiabatically compress the gas to an overdensity which is the same as that of the dark matter.

With this assumption we can approximate the Jeans mass of the gas for a fluid composed of collisionless dark matter of density, $\rho_{DM}$, and collisional baryonic material (gas) of density, $\rho_g$, and temperature, $T$, by
\begin{equation}
M_J \simeq \frac{\rho_g}{2}\left(\frac{1}{\G \rho_M} \frac{\boltz T}{\mu m_H} \right)^{\fraction{3}{2}},
\label{eq.JeansMass}
\end{equation}
where the total mass density is given by $\rho_M =
\rho_{DM}+\rho_g$.

\subsection{The Minimum Temperature for Pressure Support}
\label{sec.Intro.Tmin}
The effective mass resolution of a simulation is usually fixed.  For
simulations such as those described here which use smoothed particle
hydrodynamics (SPH) to model the gas forces, this can be taken to be
the total mass of the number of particles, $N_{SPH}$, over which local
quantities are averaged, presuming all gas particles have equal mass.
The number of particles is typically on the order of 30.  By
associating the mass of this number of particles to the Jeans mass
(\eq{eq.JeansMass}) and assuming a total mass density, $\rho_M$ in
\eq{eq.JeansMass}, equal to the mean density of the simulation volume,
$\overline{\rho}_M$, the temperature required to support gas in dark
matter halos of this size is 
\begin{equation}
T_{min} = \left( \frac{2 N_{SPH} m_g}{\overline{\rho}_g}\right)^{\twothirds} \G \overline{\rho}_M \frac{\mu m_H}{\boltz},
\label{eq.Tmin}
\end{equation}
where $m_g$ is the mass per gas particle. If the gas has a minimum
temperature set by this value, then we can say that the simulation is
resolving the Jeans mass. This assumption is flawed in a number of
ways.  Since local density is not in general equal to the mean
in the
simulation, the minimum temperature should also be a function
of position.  However, the minimum temperature should vary with the
local density as $T_{min} \simeq \rho_M^{\onethird}$, which is not a
particularly sensitive function, especially in those regions in which
the first objects are forming. More significant is the decoupling of
the components of the fluid, discussed previously.  This is likely to
occur for the first objects formed since the dark matter halos
collapse regardless of whether or not the gas is pressure-supported.
For those situations in which, locally, $\rho_{DM}\gg\rho_g$,
enhancement of the dark matter density without a similar increase in
the gas density requires the temperature to rise by an equivalent
fraction in order to maintain a constant Jeans mass for the gas.
However, during the formation of these dark matter structures, the gas
will be allowed to gradually respond, adiabatically, to the increasing
gravitational potential, partially or fully offsetting the need for an
increased minimum temperature.  This would not occur without the
initial minimum temperature since the gas in that case would initially
collapse along with the dark matter. 

\subsection{The Hypothesis}
In this paper, we propose that the effect on the state of the gas of
poorly resolving the first baryonic structures does not propagate into
larger structures formed through hierarchical merger.  The properties
of the gas in the larger halos will be determined primarily by the
virial properties of the structures which are dominated by the
distribution of the collisionless component. 

To test this hypothesis, we performed a series of seven simulations of
hierarchical structure formation.  The simulations were done at three different
resolutions.  For two of these simulations, the temperature of the gas
content was prevented from being less than the temperature required to
maintain pressure support for the smallest resolvable object; that is,
the Jeans mass was resolved.

Comparison of the data simulated for this study was made using both
the total sample of gas in each simulation volume as well as an
individual well-resolved structure.  Using an individual cluster
permits examination of those effects that are frequently
the focus of interest in structure formation simulations; the void
particles in simulations of this sort are usually ignored due to a
lack of resolution in this regime. 
 
We examine the entropy of the gas, which is a measure of the degree of 
shocking.  Also examined is the density distribution of the gas to
determine the extent of any bulk movements of the gas to higher
densities as resolution is increased.  
We also explore mean bulk properties of the
structures:  in particular, the mean dark matter and gas density
profiles and the mean baryon fraction profile.
The mean gas density and baryon
fraction profiles are potentially sensitive to a lack of convergence
as well as being of fundamental interest.

The outline of the paper is as follows.
The simulations are described in
\sect{sec.Simulations}.  The results of an attempt to recreate the
results of \citeauthor{OV97} but using a three dimensional geometry are
given in \sect{sec.Results.early}.  The impact of resolving
the Jeans mass on the final state of the gas in galaxy clusters
is addressed in \sect{sec.Results.late}.  
The results are discussed in \sect{sec.Conclusions}.
Hereafter, we refer to \citet{OV97} as
OV97.

Though the simulations are scale invariant, the results on occasion are given in physical units.  This is simply a matter of convenience for the purpose of comparison.

\section{THE SIMULATIONS}
\label{sec.Simulations}
The N-body code with SPH, \Hydra\ \citep{CTP}, was used for all
simulations.  \Hydra\ calculates the gravitational forces using an
adaptive mesh for the large-scale complemented with particle-particle
calculations for the short range forces.  Smoothed particle
hydrodynamics (SPH) is used for the hydrodynamics.  

Seven simulations were performed.  Five used the original code, and
two used a variant of \Hydra\ which imposes a minimum temperature on
the gas particles.  The details of the simulations are outlined in
\tab{tab.Sim.simulations}. Given, there, are the number of each type of particle and the gravitational softening length, $\epsilon$.  The simulations that were evolved to the final expansion factor, $a=a_f$, were performed twice, once with and once without a minimum temperature, whereas the first trio were evolved only once, without a minimum temperature.  Initial density perturbations are the same for all resolutions within a group, albeit limited by the Nyquist frequency.

The minimum temperature is set to support a mass
of particles corresponding to the effective mass resolution of the
simulation as described in the introduction.  
At each epoch of the simulation, structures composed of $N_{SPH}$ gas particles with a local density equal to the mean gas density of the volume, $\overline{\rho}_g$, and individual masses of $m_g$ are supported against collapse by the imposition of a minimum temperature, $T_{min}$.
The minimum temperature is derived by simply equating the Jeans mass, $M_J$, to the structure mass $N_{SPH} m_g$, as is done in \sect{sec.Intro.Tmin} to derive \eq{eq.Tmin}.

During the two simulations in which a minimum temperature was imposed,
$T_{min}$ was recalculated at each epoch.  Since the minimum
temperature scales with the mean density as $T_{min} \propto
\overline{\rho}^{1/3}$, then it scales with the expansion factor, $a$,
as $T_{min} \propto a^{-1}$.  Normally, adiabatic cooling  for a
uniform monotonic gas would maintain the relation, $T \propto a^{-2}$.
Consequently, the imposition of the minimum temperature has the effect
of continuously injecting thermal energy into the fraction of the gas
whose adiabatic expansion would bring its temperature below the
minimum temperature.  

The initial conditions for the two-dimensional simulations of OV97
have a density perturbation spectrum with a spectral index equivalent
to the three-dimensional $n=-1$ form used in the simulations presented
here. Because the simulations are scale free, the spatial
variance of the density fluctuations at the end of the simulation in
the larger volume corresponds to an earlier epoch in the smaller
volume, after scaling by the box size. 

The amplitude of linear density fluctuations, as measured by the power
spectrum of the density field, $P(k)$, relates to the wavenumber
of the fluctuation, $k$, measured in comoving coordinates, and the
time, $t$, via 
\begin{equation}
P(k) \propto \left(\frac{\sigma_L}{h}\right)^2 k^n t^{\fraction{4}{3}}
\label{eq.power}
\end{equation}
where $\sigma_L$ is a normalization factor corresponding to the
assumed value of the variance, $\sigma$, of the present density field
on the scale of $L$.

Although the simulations discussed in OV97 and those presented here are
scale invariant, it is useful to set a physical size to the simulation
volume. Indeed, this may be a benefit
to those readers familiar with the scale of cosmological structures.
Further, this is the only practical way we have of determining the
degree of clustering in the simulations of OV97.
OV97 uses a volume with a side of length $128
h^{-1}\Mpc$, $h=0.5$ \citep{OV97_preprint}. Since we use a volume with
sides $40 h^{-1}\Mpc$ and $h=0.65$, this gives a ratio of scales of
$4.16$ for their simulation scale compared to the simulations
described here. For the normalization factor, they have $\sigma_8=0.6$
whereas we used $\sigma_8=0.935$. This implies an expansion factor of
$0.24 a_f$.  That is, the structures in the simulations presented
here, scaled by the box size, will have grown in similar magnitude to
those of OV97 after an expansion of $0.24$ of the final expansion
factor of the simulation. 

In this study, the data have been compared at the slightly earlier
expansion factors of $0.17 a_f$ and $0.22 a_f$ to ensure that
structures have not undergone more than a couple of merger events.
As a check, a volume with the same cosmological parameters as described in OV97 but with
the same perturbation waves as found in our smaller boxes was evolved
to an expansion factor, $a_f$.  The data from the the $40 h^{-1}\Mpc$
volumes at an expansion factor of $0.22 a_f$ are visibly similar to
the final output of this test volume, supporting the legitimacy of
scaling the numerical results. 

It was feasible, computationally, to evolve the volume to the present
only at the resolutions of  
$2\times 64^3$ particles and less.  Hence, this study will concentrate
on a comparison of the simulations of resolution corresponding to this
number of particles with simulations of a
resolution appropriate to $2\times 32^3$ particles. 
These resolutions will be referred to henceforth as the $64^3$ and $32^3$
simulations.  Each was evolved with and without a minimum temperature.
For the earlier epoch corresponding to the expansion factor of $0.17
a_f$, the volume could also be evolved at the resolution provided by
$2\times 128^3$ particles.  Thus a set of three simulations was
performed using the same density perturbations (limited by the Nyquist
frequency) at resolutions corresponding to $128^3$, $64^3$, and
$32^3$. 

Cooling was neglected to maintain the scale invariance.  In all simulations, $\Omega=1$, $\Omega_{DM}=0.9$, and $\Omega_g=0.1$.

The initial density perturbations were established by displacing the
particle positions from a uniform cubic grid using
the \citet{Zeld70} approximation for growth of density perturbations
in the linear regime in the standard way. 

The initial redshift for the $32^3$ and $64^3$ simulations is $z_{init}=75$. This was chosen to keep the maximum displacement incurred during the establishment of the density field to less than \onehalf\ of the initial grid spacing.  This ensures that the correct non-linear field will be attained at later times.  The same criterion was used to set the initial redshift for the $128^3$ simulation for which $z_{init}=150$. 

\section{RESULTS}
\label{sec.Results}
\subsection{The State of the Baryons at an Early Epoch}
\label{sec.Results.early}
We first attempt to reproduce the results of OV97 by examining the
state of the baryons in our simulations at a comparable epoch with
that of the final epoch of the simulations of OV97.  
The smaller physical scale of the simulations presented
here requires that we examine our data before the nominal end of the
simulation 
in order that a comparable level of clustering has developed to that
present  at the final
epoch of the simulations of OV97.  The required expansion factor is
$a=0.22a_f$, where $a_f$ is the final expansion factor of the
simulations.  

Following the analysis given by OV97, we look at the distribution of
the gas particles in the $\rho-T$ plane as well as the
density distributions for which the density is resampled down to the
resolution of the lowest resolution runs as described below.  The
position within the $\rho-T$ 
planes gives an indication of the degree of shocking, with the more
highly shocked gas (gas with greater entropy as indicated by $T
\rho^{-\gamma}$, $\gamma=\twothirds$) lying to the upper left (high
temperature, low density). 

In order to allow a direct comparison of the densities in simulations
of different resolution a density resampling technique was employed.
The (resolution-degrading) resampling was accomplished by smoothing
over equal masses for each particle using an SPH-like density
estimate.  Consequently, there are approximately eight times as many
particles in each estimate for the $64^3$ data as in the $32^3$ data,
but the smoothing radii are approximately the same. This resolution
degradation is performed for the high resolution simulations ($64^3$)
when comparison is made with the lowest resolution simulation
($32^3$). This technique was
unfortunately not possible with the $128^3$ data-set due to the
computational effort of summing over the more than 3000 neighbor
particles of each gas particle.

It is clear that for density
profiles without cores, higher mass and spatial resolutions will lead to
more gas at a higher density even if the simulations can be said to
have converged to the true solution within resolution limits. If the
densities are averaged within volumes encompassing a mass which is
resolved in the lowest resolution simulation, then these masses should
not vary significantly.  

The simulations of OV97 produce more cool, dense gas as the resolution
increases.  They interpret this to mean that in the higher resolution
simulations, the gas in the cores of structures is not as strongly
shocked.  
To search for this effect, three sets of data were
evolved at the resolutions provided by $2\times32^3$, $2\times64^3$,
and $2\times128^3$ particles, half in each phase of gas and dark
matter, as described in \sect{sec.Simulations}.  Each set has the same
initial perturbations, limited appropriately in the high spatial
frequency domain by the Nyquist frequency. These data sets were
evolved to an early epoch for which the amplitude of density
fluctuations has grown to a level comparable to those in the data
presented by OV97. 

The evolved data were resampled, in the case of the $64^3$ and $128^3$
sets, to the same mass resolution as the $32^3$ set by randomly
selecting $32^3$ particles from each phase.  This avoids the problem
of performing the density estimate over the very large number of
neighbor particles required for the $128^3$ data set, at the expense
of an increase in shot noise.  Local SPH density estimates for the
sampled particles in these resampled sets were determined allowing
comparisons of the data in $\rho$-$T$ space. 

\begin{figure}
\plotone{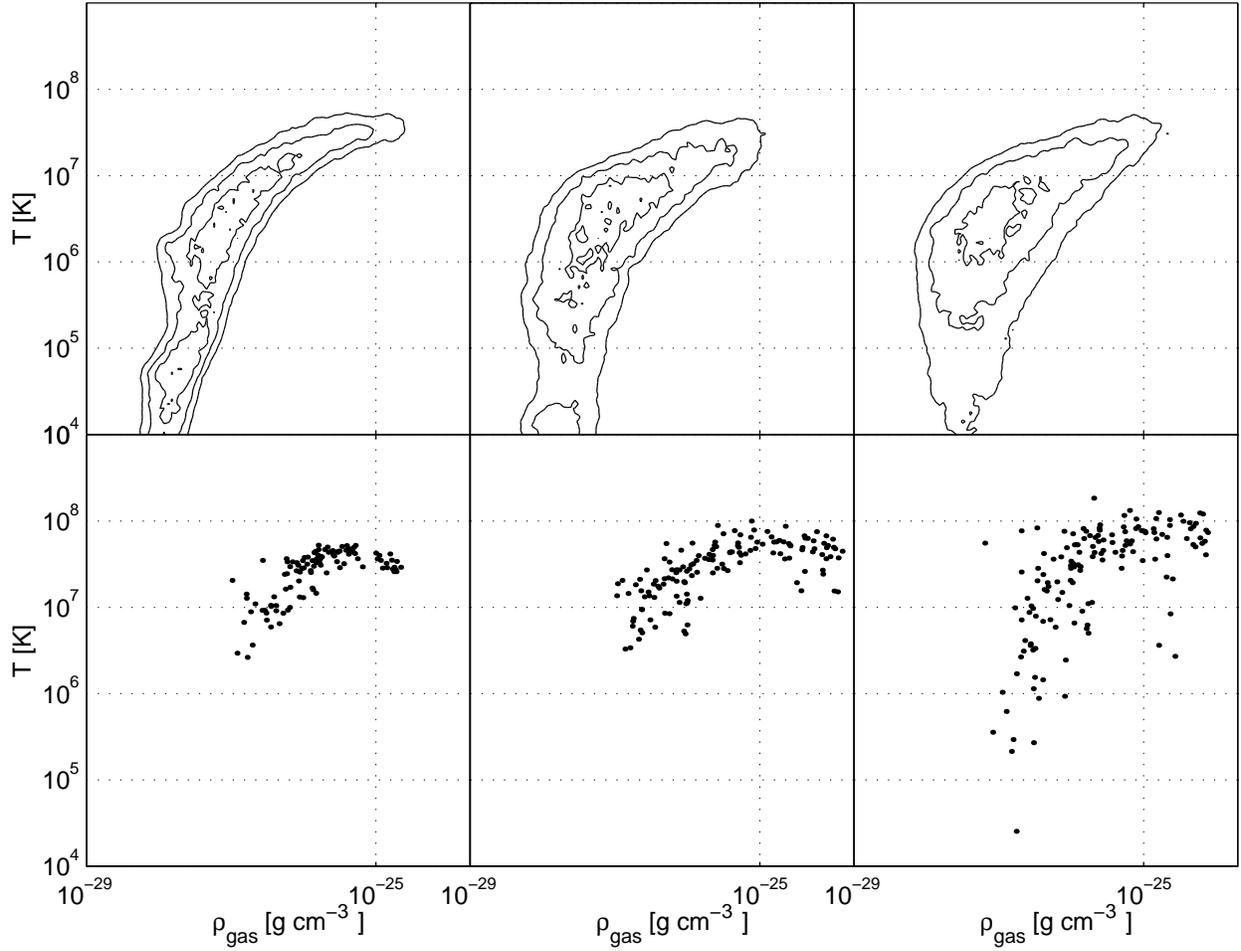}
\caption{Temperature {\it vs.} gas density at a timescale factor, $a=0.17 a_f$.  The upper set of panels are for all the gas while the lower set is for a selected cluster.  From left to right, the panels correspond to simulations with increasing resolutions of $32^3$, $64^3$, and $128^3$ baryon particles.  The simulation data were resampled before analysis to the same mass resolution as the $32^3$ simulation. Contour lines in the upper panels are of constant mass surface density in $\rho-T$ space.  In the lower set of panels, the data points are for the sample of particles in the selected cluster.}
\label{fig.Results.TvsRho_3Res}
\end{figure}

The $\rho$-$T$ distributions for all the gas in each of these three
simulations are illustrated in \fig{fig.Results.TvsRho_3Res} (top
panels).  This is comparable to the top panels of Fig.~8 of OV97. It
is clear that there is no trend for the {\em bulk of the mass} to move
toward a cooler and denser state as the resolution increases, in
contrast to the results of OV97.  Since the bulk of this gas is in the
halos of bound structures, the argument given in the introduction is
supported; the entropy of the gas is set by the virial properties of
the halo which are dictated by the dark matter.  There
is a trend for the void particles (found to the lower left in the
$\rho$-$T$ plane) to increase in temperature as the resolution
increases. There is also a trend for the distributions to become more
diffuse, particularly toward the denser and cooler state. 

The lower set of panels in \fig{fig.Results.TvsRho_3Res} give the distribution for particles in a cluster common to all three sets of data.  The cluster was selected simply for being one of the larger clusters of particles to have formed by this early stage.  The cluster is barely resolved ($\sim 100$ particles) in the $32^3$ simulation, but is well resolved in the $128^3$ simulation.  Here it is evident that the diffusion of the distributions to a denser and cooler state is due to a small population of particles, and not a bulk movement. Upon examination of the spatial distribution of this population, these particles are found to be located in substructure that is present only in the simulations with greater mass resolution.  These clumps are not as shocked as the particles in the halos of the substructure.  This is consistent with the phenomenon described by OV97 in nature, though not in extent since it is not a bulk motion, but simply a new population of particles.

\begin{figure}
\epsscale{\figscaletwo}
\plotone{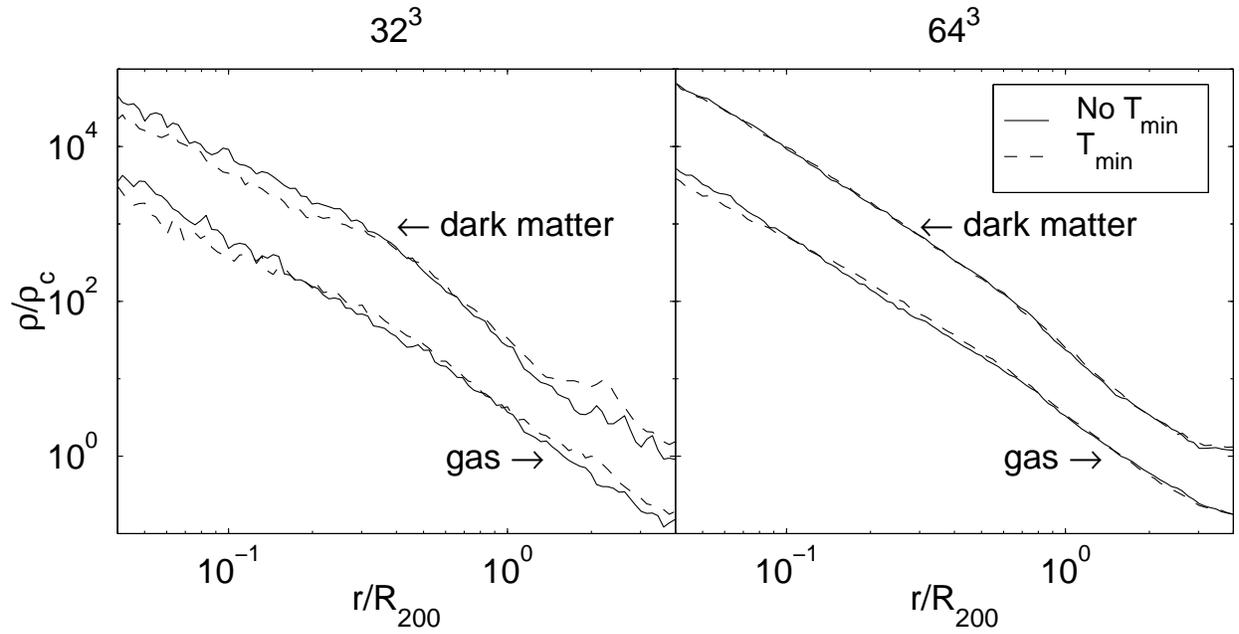}
\caption{The mean dark matter and gas density profiles.  The panel to
the left corresponds to the low resolution set of runs, while the
panel on the right is for the high resolution set.  Profiles are given
for clusters formed in the absence of a minimum temperature and for
clusters formed in a simulation with a
minimum temperature scaled appropriately for the mean gas density at
each epoch.} 
\label{fig.Results.RhoProfiles}
\end{figure}

As the mass resolution increases, the first bound haloes form at
earlier epochs at higher densities leading to increased central
densities in the final dark matter halos. This is illustrated in \fig{fig.Results.RhoProfiles} which compares the gas and dark matter mean density profiles for resolved clusters in the simulations, albeit at a later epoch.  The dark matter density is double in the $64^3$ simulation compard with the $32^3$ simulation in the central regions (close to the resolution limit of the $32^3$ simulation). This {\em increases} the degree of shocking of the halo particles more than offsetting the decreased amount of shocking in the cores of substructure.  The mean temperature of this structure thus {\em increases} with increasing resolution: in physical units it is $2.9\expd{7}\K$, $3.2\expd{7}\K$, $4.5\expd{7}\K$ for the $32^3$, $64^3$, and $128^3$ simulations, respectively.  The total amount of shocking on the whole also increases, but not to such a degree.  The amount of shocking, as measured by the mass-weighted mean of the entropy parameter, $<T/\rho_g^{\twothirds}>_M$, for an individual cluster at the three resolutions evolved to $a=0.17 a_f$ is given in \tab{tab.Results.T_3Res}.  The values not resolution-degraded are given in the second column which reveal a slight increase (10\% for every doubling in resolution) in the entropy as the resolution increases.  If the densities are recalculated for the resampled data with the intent of comparing similar effective resolutions, then the degree of shocking decreases quite markedly with increasing resolution (\tab{tab.Results.T_3Res}, third column). This occurs simply because the temperatures of the particles are not changing while the resampled densities of the particles in the densest regions decrease due to the use of a larger smoothing radius.  It would be erroneous to conclude that the gas is actually less shocked, since the temperature of the halo gas particles and the depth of the potential well are so intimately related. 

\begin{figure}
\epsscale{\figscale}
\plotone{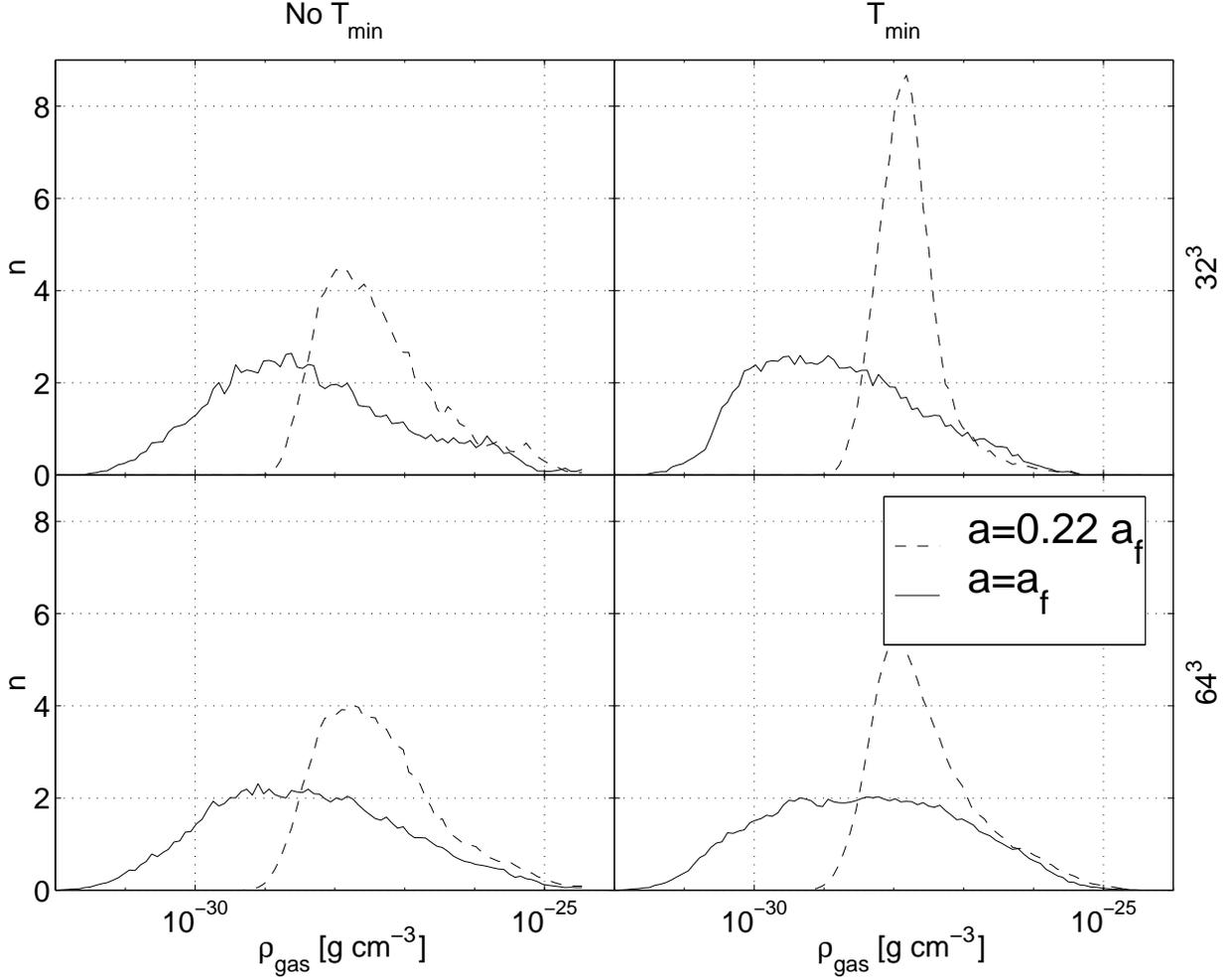}
\caption{The distribution of gas density at early and late epochs. The upper set of panels is for the $32^3$ set of data while the lower set is for the $64^3$ set.  From left to right, the panels correspond to no minimum temperature and a minimum temperature scaled appropriately for the epoch.}
\label{fig.Results.RhoDist}
\end{figure}

The gas density distributions for both the $32^3$ and
$64^3$ data were found at an expansion factor of $0.22 a_f$. 
The densities of the $64^3$ data-set were resampled as described
previously. The distribution of gas density 
at this early epoch (\fig{fig.Results.RhoDist}, left hand panels,
dashed lines) is only weakly dependent on resolution, becoming
somewhat less peaked at $10^{-28}\g \cm^{-3}$ at the higher
resolutions.  No bulk motion of the gas to higher densities is
observed, unlike the simulations described in OV97. 

\begin{figure}
\plotone{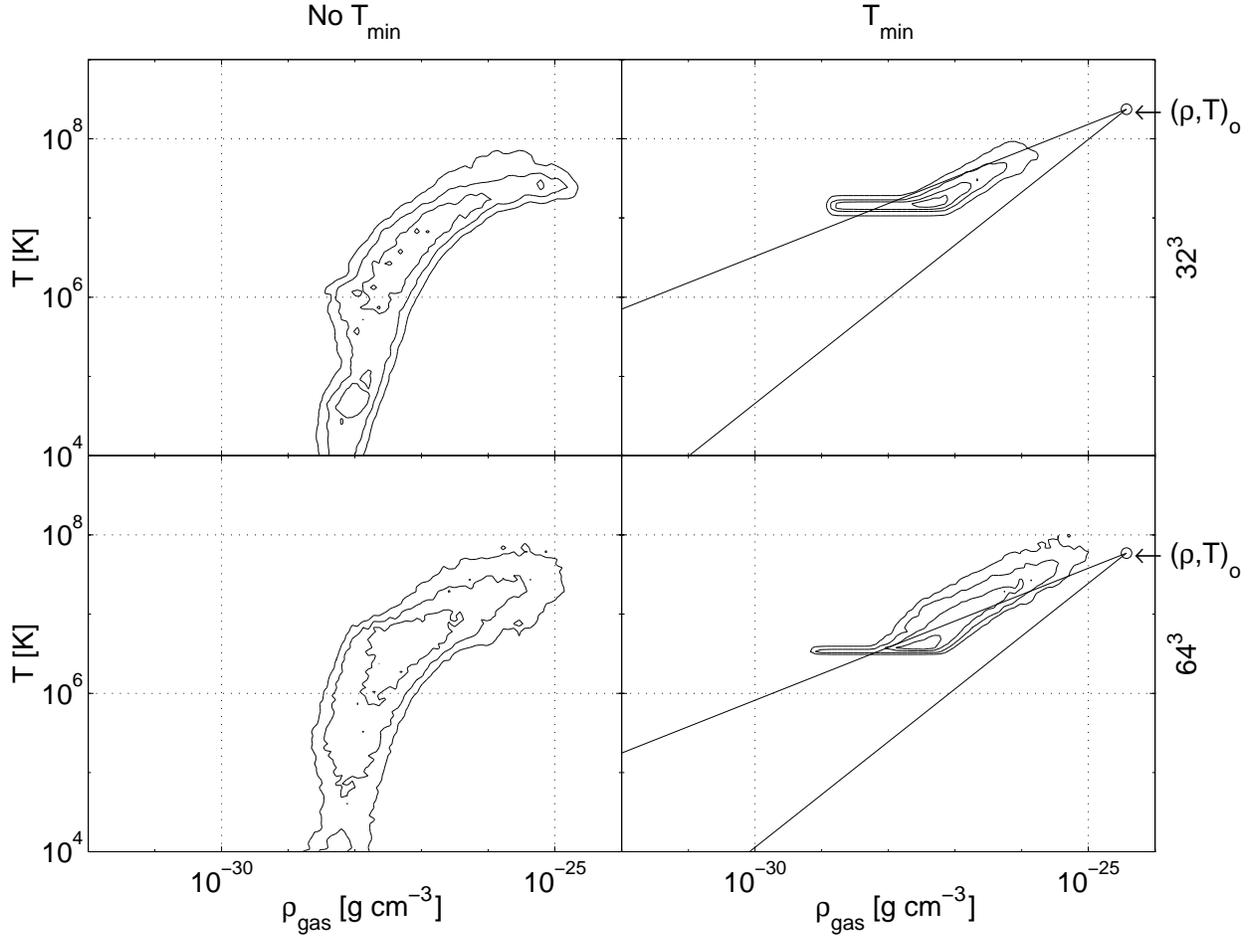}
\caption{Temperature {\it vs.} gas density for all the gas at an
expansion factor of $a=0.22 a_f$.  The circles in the right-hand panels
indicate the starting density and temperature.  For the left-hand
panels, these points are off the figure at $T=100\K$. The lines projecting
to the lower left of each circle correspond to the path of taken by a
uniform gas expanding and cooling adiabatically (lower line) and to
the path of gas
maintaining $T_{min}$ (upper line).  The panels are arranged as in
\fig{fig.Results.RhoDist}.} 
\label{fig.Results.TvsRho_0.1}
\end{figure}

Since no bulk motion of the gas to higher densities and lower degrees
of shocking is observed, we might predict that the imposition of a
minimum temperature will not significantly affect these results. 
The distribution of the gas within the $\rho-T$
plane at the early epoch of $a=0.22a_f$ is illustrated in
\fig{fig.Results.TvsRho_0.1}.  Again, the bulk of the gas is found in
the same state in both simulations without a minimum temperature
(\fig{fig.Results.TvsRho_0.1}, left panels), with a diffusion toward a
state of lower entropy.  For those simulations in which a minimum
temperature is imposed (\fig{fig.Results.TvsRho_0.1}, right panels),
the gas is displaced within the $\rho-T$ plane in those regions below
$T_{min}$ (for obvious reasons) and below the locus that uniform gas
would trace if it were allowed to cool to the minimum temperature
appropriate at each epoch.  
In order for gas to be found below this locus it must be initially
held at a high temperature and density but with little or no increase in entropy
and subsequently allowed to cool adiabatically.
The bulk of the gas is above these boundaries.  The gas
that is above these boundaries is consistently located in similar
proportions to the gas found in the same locations in the $\rho-T$
plane in the simulations without a minimum temperature.  There is no
gas in the lower resolution simulations which fits these criteria,
however.
  
The resampled density distributions for the simulations with a minimum temperature, illustrated in \fig{fig.Results.RhoDist} (right panels, dashed lines), confirm that the minimum temperature does modify the distributions but does not provide convergence.  This figure is comparable to Fig.~6d in OV97 which indicates a bulk motion to higher densities as the resolution is increased, even when the data are resampled to the scale of the lowest resolution simulation.  In our simulations, for the gas denser than $10^{-27} \g \cm^{-3}$, there is actually a decrease in the agreement between the lower and higher resolution data when a minimum temperature is enforced.

The decrease in the degree of shocking reported by OV97 as the resolution of a hydrodynamic simulation is increased is confirmed in the cores of clusters of particles.  However, the bulk movement of the gas particles to a less shocked state is {\em not} observed in our simulations.  Indeed, the mean degree of shocking of cluster particles is essentially conserved, with the mean temperature of particles actually increasing with resolution.  The displacement in density distributions reported by OV97 is also not observed, nor is any sort of convergence provided by the imposition of a minimum temperature.

\subsection{The State of the Baryons at a Late Epoch}
\label{sec.Results.late}
After many mergers, it is possible that the gas in the initial structures described above will have `forgotten' the degree of shocking it had at early times.  It may have been sufficiently shocked by subsequent mergers that early heating, via strong shocks or otherwise, would no longer be significant to the total thermal content and distribution.  To examine this hypothesis, the $32^3$ and $64^3$ sets of simulations used in \sect{sec.Results.early} were evolved to a further state which involved many levels of merging.  It is not presently feasible to evolve a $128^3$ to such a state with the hardware on hand in a manageable amount of time.

\begin{figure}
\plotone{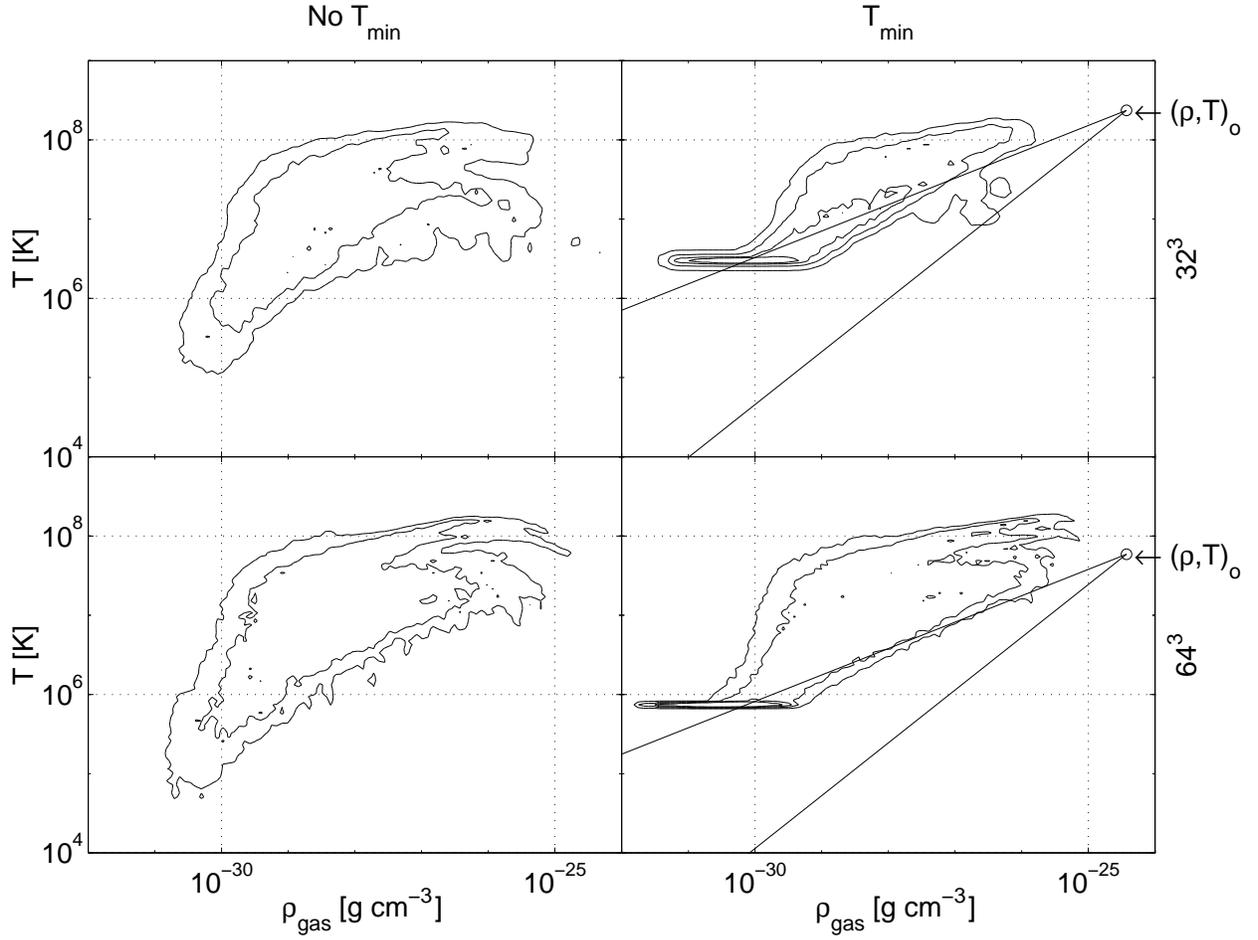}
\caption{Temperature {\it vs.} gas density for all the gas at an expansion factor of $a=a_f$.
The panels are as described in \fig{fig.Results.TvsRho_0.1}.}
\label{fig.Results.TvsRho_1}
\end{figure}

The gas distributions in the $\rho$-$T$ plane at this end state are illustrated in \fig{fig.Results.TvsRho_1}.  At densities less than $\sim 5\expd{-26}\g \cm^{-3}$, the distributions for the $32^3$ (top left) and $64^3$ (bottom left) simulations are identical in form.  At $\sim 5\expd{-26}\g \cm^{-3}$, it is the lower mass-resolution simulation that produces the greatest amount of cool, dense gas.

Evolving these sets of data while enforcing a minimum temperature
sufficient to pressure support structures with $N_{SPH}$ or fewer
particles does little to alter these distributions
(\fig{fig.Results.TvsRho_1}, panels to the right) outside those
regions below $T_{min}$ and below the line with a slope of \onethird\
which extends from the point of initial density and temperature of the
gas.  This line corresponds to the path a uniform gas would take due
to the imposition of a minimum temperature. 

\begin{figure}
\plotone{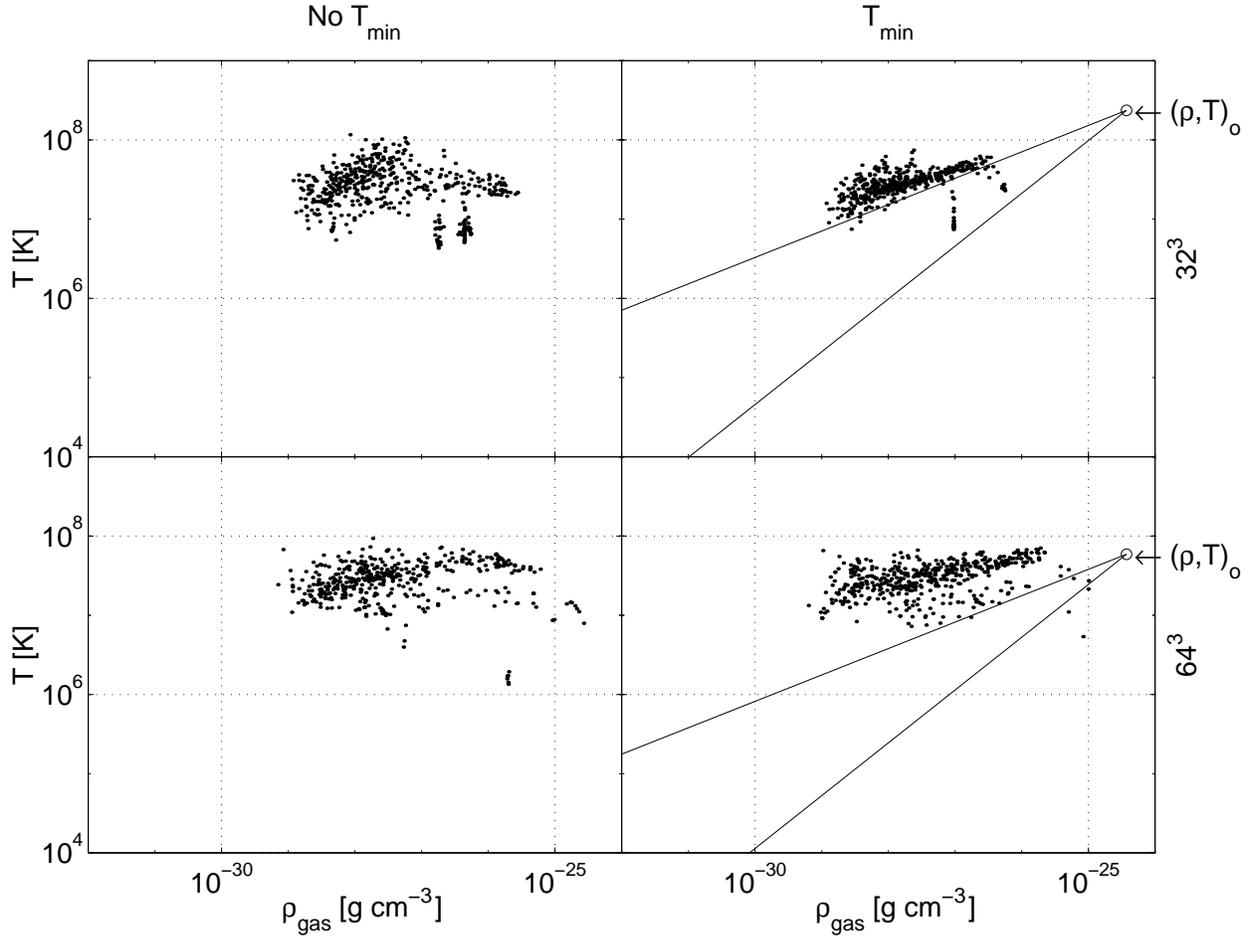}
\caption{Temperature {\it vs.} gas density for the gas of a single cluster at an expansion factor, $a=a_f$.
The panels are as described in \fig{fig.Results.TvsRho_0.1}.}
\label{fig.Results.TvsRho_cluster_1}
\end{figure}

Recall from \sect{sec.Results.early} that individual structures in
high resolution simulations have, at early times, less shocked
material in their cores than is produced at lower mass
resolution. This difference is eliminated after the initial merging period on
account of the introduction of new structures falling into the
lower-resolution cluster. These structures have material with low
degrees of shocking. Compare the lower panels of
\fig{fig.Results.TvsRho_3Res} with \fig{fig.Results.TvsRho_cluster_1}
(left panels) which plots the $\rho$-$T$ distribution for a cluster
from a $32^3$ and $64^3$ simulation at late times (note that the
figures have different scales).  At early times, there is a
population of less shocked gas in the high resolution runs as described
earlier.  This leads to the spread in temperature at high densities.
Although the distribution of substructures differs among the simulations
of differing resolution, the presence of substructures at the termination of the simulations leads to a similar spread in temperatures, regardless of resolution. 
Taken as singular objects, the bulk properties of individual clusters
do not vary 
substantially between simulations of differing resolutions.  The
degree of shocking, as parameterized by the entropy factor
$<T\rho^{-\gamma}>_M$ (a mass-weighted mean) is found to be
essentially constant for the cluster simulated at either resolution.
It is $(1.12\pm 0.02)\expd{26} \K\cm^2\g^{-\twothirds}$ for both the
$32^3$ and $64^3$ simulations.  With the imposition of a minimum
temperature, the entropy is essentially unchanged at
$(1.13\pm 0.03)\expd{26} \K\cm^2\g^{-\twothirds}$.  The trend for the
mean temperature of structures to increase with increasing resolution
is maintained, with the gas temperature rising from $(2.9\pm
0.1)\expd{7}\K$ to 
$(3.2\pm 0.1)\expd{7}\K$ between the $32^3$ and $64^3$ simulations.
This is understood to be a by-product of the deeper potential wells in
the cores of structures as described earlier. 

The density distribution confirms the similarity of the states of the
gas in the simulations of differing resolution
(\fig{fig.Results.RhoDist}, solid line).  In the absence of a minimum
temperature (\fig{fig.Results.RhoDist}, left panels), the gas density
distribution varies marginally with resolution with a decrease in the
amount of gas at a density of $\sim 10^{-29} \g \cm^{-3}$ compensated
by an increase at the higher density of $\sim 10^{-27} \g \cm^{-3}$ as
the spatial resolution is doubled.  The addition of a minimum
temperature skews the distribution even more significantly to lower
densities for the lower-resolution simulation.  The shift in the
simulations without a minimum temperature can be explained as a result
of the greater abundance of substructure in the dark matter
distribution which locally compresses and heats the gas.  The gas at
the highest densities in the cores of structures ($\rho>10^{-26} \g
\cm^{-3}$), is not significantly affected. 

Finally, a useful diagnostic is the mean density profile (\fig{fig.Results.RhoProfiles}) of the clusters at the final epoch, $a=a_f$.  It confirms that the presence of a minimum temperature does not change the morphology of the clusters.  The mean profiles are calculated from a sample of clusters selected by their size, to ensure that the profiles are resolved.  The inner cores are excluded from the calculation of the means, since they are not resolved.

\begin{figure}
\epsscale{\figscale}
\plotone{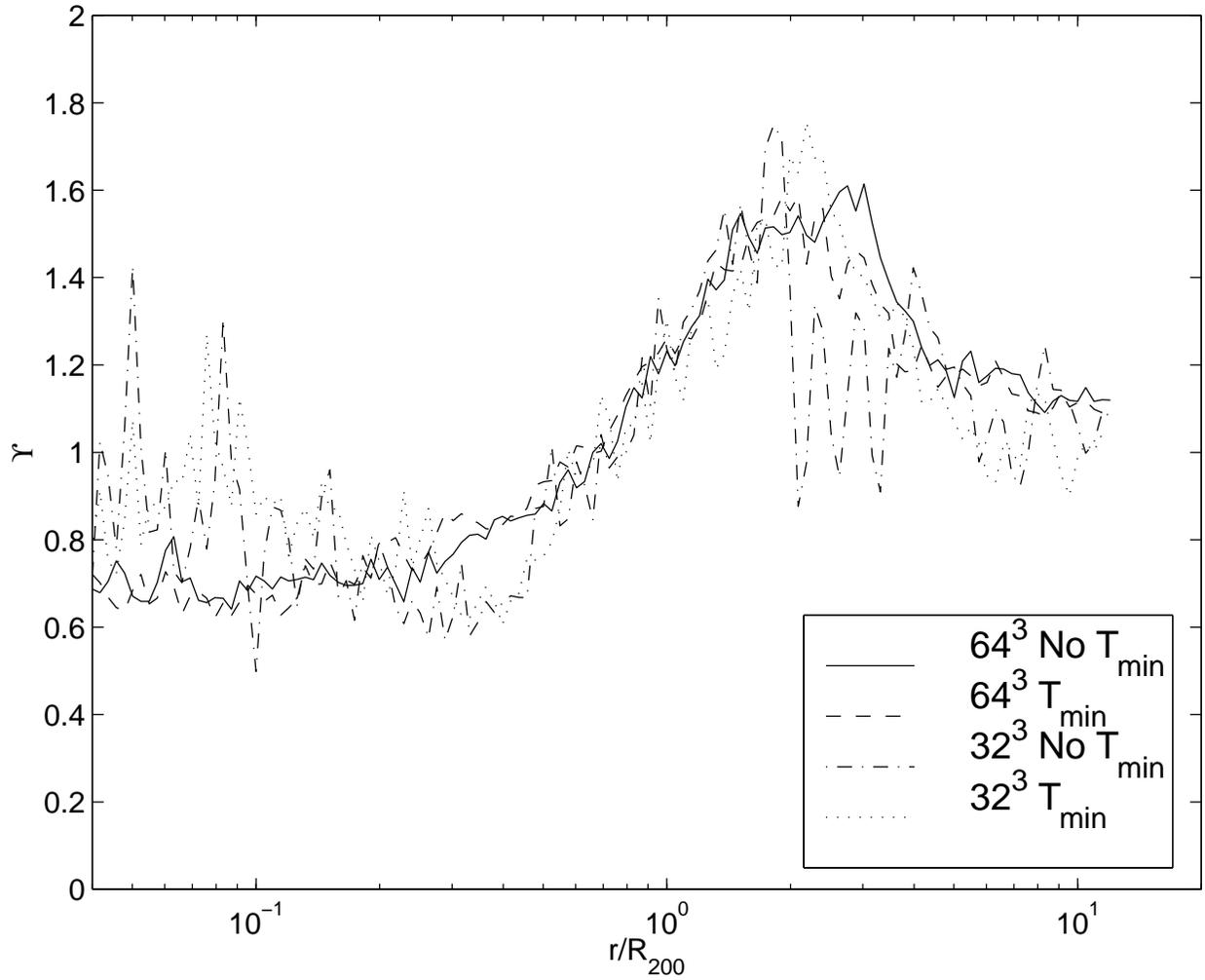}
\caption{The radial profile of the mean baryon fraction parameter, $\Upsilon$, for the clusters formed at different resolutions in the presence or absence of a minimum temperature. }
\label{fig.Results.BFProfiles}
\end{figure}

The simulations of OV97 display a marked tendency for the gas to become enriched in the cores of clusters, illustrated in Fig.~9 of OV97.  This is contrary to the results found by others \citep{Evrard90,TC92,CO93a,Kang94,ME94,PTC94,NFW95,AN96,Lubin96,PEG96}.  Indeed, it is most commonly found that the gas becomes more dispersed than the dark matter.  This result is supported by the simulations we performed.  The baryon fraction parameter, $\Upsilon$, which is the ratio of the baryon density to total mass density normalized by the global mean ratio, is plotted in \fig{fig.Results.BFProfiles} for the clusters.  The profiles given are the mean profiles for the same sample of clusters for which the mean density profiles have been calculated.  It is also clear from this figure that the presence of a minimum temperature has no effect on the baryon fraction distribution.  It also shows clearly that resolution does have an effect in the cores.

\section{DISCUSSION}
\label{sec.Conclusions}
It has been suggested \citep{OV97} that convergence with increasing
mass resolution in numerical simulations of hierarchical structure
formation requires the imposition of a minimum temperature in order
that gas is pressure supported in structures with masses similar to
the effective resolution of the simulation. Our results do not support
this contention, either at early epochs after only a few mergers, or
at later epochs after a significant number of mergers have occurred
and the minimum temperature has dropped well below the virial
temperature of the typical structures.
Our results indicate that the problem of the gas becoming over-shocked
in simulations in which the Jeans Mass is not resolved is not as
severe as suggested by OV97. The larger degree of shocking of the bulk
of the gas noted by OV97 as resolution is lowered is not reproduced at
any epoch of the simulations presented here.  Consequently, the
imposition of a minimum temperature is not necessary to achieve
convergence. 

We do observe a phenomenon which is similar to the decrease in
shocking with increasing resolution as described in OV97. At an
early epoch, isolated halos are shown to possess a large amount of
substructure in their high resolution simulations that is not present
in their low resolution counterparts. This substructure contains
cool, dense, weakly shocked gas.  However, the presence of this
population of cool and dense gas in the high resolution simulations is
offset by an increase in the temperature of the halo gas surrounding
these substructures.  The net effect is not to lower the mean
temperature of the halo but to marginally increase it ($10\%$ for each
doubling in resolution) as well as to increase the entropy of the halo
by an
equivalent fraction.  At a later epoch, after subsequent accumulation
of material, the halos in the lower resolution simulations develop a
comparable population of cooler, unshocked gas within substructure.

Further attempts were made to reproduce the results of OV97 by
modifying our code.  Since it was
believed that the gravity calculations for the simulations of OV97
were done using a particle-mesh method, we disabled the
particle-particle contributions of the gravitational force as well as
the refinements in the \Hydra\ code.  This was unsuccessful in
reproducing their results. 

The results of imposing a minimum temperature sufficient to pressure-support the resolution-limited gas structures were examined for adverse effects.  At the final epoch, the gas density profiles of the clusters were not altered.  The degree of shocking increased by less than $5\%$.  This increase may be attributed primarily to an increase in temperature and decrease in density of the gas lying near the cores of structures.  The bulk of the gas, which lies in the halo, was unaffected.

The concentration of baryonic material in the cores of structures
reported by OV97 is also not observed. Instead, an anti-bias is found,
in accordance with the substantial majority of numerical simulations
reported in the literature as noted previously.
\citet{PTC94} explains the phenomenon as a result of the merging
process of gas-dark-matter halo systems in which gas is shocked,
permanently removing the energy from the dark matter component.  For
the gas to be more biased in clusters, without an energy-loss
mechanism operating such as cooling, the gas must transfer energy to
the dark matter component.

It is unclear what causes the differences between our results and
those of OV97. The most striking remaining difference between our
investigations and those of OV97, which we have not explored, is the
difference in dimensionality of the numerical simulations.  OV97 used
two-dimensional simulations, ours were three-dimensional.  Structures
in two dimensions are rods, which have essentially the same cross
sections per mass as spheres, but one less degree of freedom in which
to avoid collisions.  This may strongly affect the efficiency of shock
heating, with more head-on collisions per merging event leading to
more violently shocked gas.  However, na\"{i}ve interpretation of this
observation would suggest that the gas halos should be more extended,
and more anti-biased relative to the dark matter, contrary to their
results.

We conclude that resolution of the Jeans mass of the smallest
structures is consequently not important in studies of the gas in
structures after merging has occurred.  It may be important for the
first objects formed, but for any study other than the statistics of
position and number, these first objects should be ignored since they
are, by definition, poorly resolved.  Clearly, this result may not hold for
simulations in which there is cooling and/or star formation feedback,
but these will have an effect only in the densest parts of the
clusters, and should not affect  global properties. 

\acknowledgements

We thank Rob Thacker for helpful discussions. HMPC acknowledges the
support of NSERC.

\bibliographystyle{apj}
\bibliography{apj-jour,biblio}

\renewcommand{\thisdir}{Tables}

\begin{deluxetable}{cccc}
\tablecolumns{4} 
\tablewidth{0pc} 
\tablecaption{Simulation properties}
\tablehead{
\colhead{}  & \multicolumn{2}{c}{\# of particles} & \colhead{$\epsilon$} \\
\colhead{Resolution} & \colhead{gas} & \colhead{dark} & \colhead{($h^{-1} \kpc$)} \\
\cline{1-4}
\multicolumn{4}{c}{Evolved to $a=0.17 a_f$}
}
\startdata
$32^3$ 	& $32^3$&  $32^3$ 		     & 80            \\
$64^3$ 	& $64^3$&  $64^3$ 		     & 40            \\
$128^3$ &$128^3$& $128^3$ 		     & 20            \\
\cutinhead{Evolved to $a=0.22 a_f$ and $a_f$}
$32^3$ 	&$32^3$& $32^3$ 		     & 80            \\
$64^3$ 	&$64^3$& $64^3$ 		     & 40            \\
\enddata
\label{tab.Sim.simulations}
\end{deluxetable}

\begin{deluxetable}{rcc}
\tablecolumns{3}
\tablewidth{0pc}
\tablecaption{The measure of shocking of an individual cluster}
\tablehead{
\colhead{} & \multicolumn{2}{c}{$<T/\rho_g^{\twothirds}>_M$ ($\K \cm^2 \g^{-\twothirds}$)} \\
\colhead{Resolution} & \colhead{Unsampled} & \colhead{Sampled and recalculated $\rho_g$} 
}
\startdata
$32^3$	& $5.5\expd{24}$ & $5.5\expd{24}$ \\
$64^3$	& $5.9\expd{24}$ & $1.6\expd{24}$ \\
$128^3$	& $6.5\expd{24}$ & $0.5\expd{24}$ \\
\enddata
\label{tab.Results.T_3Res}
\end{deluxetable}

\renewcommand{\thisdir}{Figures}

\end{document}